\documentclass[onecolumn,noshowpacs,floatfix,aps,showkeywords,11pt,nofootinbib]{revtex4}
\usepackage{ucs}
\usepackage{textcomp}
\usepackage{mathtools,harmony,upgreek}
\usepackage{color,accents}
\usepackage{mathrsfs}
\usepackage{amssymb}
\usepackage{amsmath}
\usepackage{dsfont}
\usepackage{indentfirst}
\usepackage{latexsym}

\newcommand{\cl}{C \kern -0.1em \ell}

\newcommand{\be}{\begin{equation}}
\newcommand{\ee}{\end{equation}}
\newcommand{\ba}{\begin{array}}
\newcommand{\ea}{\end{array}}

\newcommand{\beq}{\begin{eqnarray}}
\newcommand{\eeq}{\end{eqnarray}}

\definecolor{clearblue}{rgb}{0,0.5,0.9}
\definecolor{orange}{rgb}{1,0.5,0}
\begin{document}

\title{Probing Topologically Charged Black Holes on Brane Worlds in $f(\mathrm{R})$ Bulk}\author{Andr\'e  M. Kuerten}\email{andre.kuerten@ufabc.edu.br}
\affiliation{ CCNH, Universidade
Federal do ABC 09210-580, Santo Andr\'e,  Brazil}
\affiliation{Instituto de Ciencias F\'{\i}sicas, Universidad
Nacional Aut\'onoma de M\'exico
MEX--62210, Cuernavaca, Morelos, M\'{e}xico}
\author{Rold\~ao da Rocha}
\email{roldao.rocha@ufabc.edu.br}
 \affiliation{CMCC, Universidade
Federal do ABC 09210-580, Santo Andr\'e, SP, Brazil}
\begin{abstract}
The perihelion precession, the deflection of light and the radar echo delay are classical tests of General Relativity here used to probe brane-world topologically charged black holes in a $f$(R) bulk. Moreover, such tests are used to constrain the parameter that arises from the Shiromizu-Maeda-Sasaki procedure applied to a $f$(R) bulk. Observational data constrain the possible values of the tidal charge parameter and the effective cosmological constant in this context. We show that the observational/experimental data for both perihelion precession and radar echo delay make the black hole parameters to be more strict than the ones for the DMPR black hole. Moreover, the deflection of light constrains the tidal charge parameter similarly as the DMPR black holes, due to a peculiarity in the equation of motion.
\end{abstract}
\maketitle
\flushbottom

\section{Introduction}

Brane-world models play a prominent role on  high energy physics, inspired in string theory advances. This framework has cosmological and astrophysical
 implications comprehensively investigated in the
literature \cite{RS99b,gog2,harko1,harko2,barbosa,daro,HoffdaSilva:2009ht,casad}. Besides, 5D effects originated from the gravitational collapse have been 
proposed \cite%
{casadioa,bruni,ca3,CoimbraAraujo:2005es}. An interesting aspect of cosmology is that the Universe goes through a phase of the accelerated expansion,  supported by recent observational data  \cite{rie}, what can be accounted for  either dark energy  or modified theories of gravity \cite{b2} as well. Although the Einstein-Hilbert action can be replaced by an arbitrary  function $f(R)$ of the 4D Ricci scalar $R$ \cite{od2}, a Randall-Sundrum type model with $f$(R) as the action in bulk space is still incipient in the literature \cite{Carames} (hereon we denote by R the 5D Ricci scalar). 
 Moreover, recently the 5D $f$(R) theories of gravity have been studied \cite{huang} to address the dark matter problem, whereas a $f$(R) model of gravity with curvature-matter coupling in a 5D bulk was 
established in \cite{wu}.
The $f$(R) framework has been further employed to solve the brane effective field equations for   dark pressure and dark radiation to acquire  black hole solutions, with parameters induced from the bulk \cite{chakra}.

On the other hand,  General Relativity (GR) explains  the deflection of light and the perihelion shift of Mercury, complying with great accuracy to the experimental/observational values   in the context of the Schwarzschild geometry. Such classical tests were further employed in the framework of brane-world gravity \cite{boemer}. 
Our aim is to study these models,  encompassing  $f$(R) bulk effects, and probe black holes derived in such a context, by using the classical tests. In fact,  the field equations on the brane have been recently  solved, obtaining a topological brane-world black hole from a  5D $f({\rm R})$ action \cite{rengifo}. Theories that take into account 4D $f(R)$ effects are natural scenarios that unify and explain both the inflationary paradigm and the dark energy problem. Hence, it is natural to go beyond and consider both the brane-world model and the modified gravity likewise. 
Brane-world models may explain the current  acceleration of our Universe, while the 4D $f(R)$ theories can either apply to the early Universe inflation or late time acceleration,  depending on specific forms chosen. Here we analyse the physical consequences of merging both frameworks.
 The geometry to be employed here, ruled by topologically charged $f$(R) brane-world black holes, is more general and is led to both DMPR and Schwarzschild-de Sitter solutions for suitable limits of parameters. Randall-Sundrum like models,  with $f$(R) as the action in bulk,  were presented in \cite{sepangi} by using a generalized Shiromizu-Maeda-Sasaki procedure \cite{SMS00}. Nevertheless, there is a quantity $Q_{\mu\nu}$  originated in the geometry of the bulk by the function $f$(R) that describes matter \cite{rengifo}. Since  $Q_{\mu\nu}$ appears in the metric of topologically charged $f$(R) brane-world black holes, we aim  to study it by the classical tests of GR. Thus, it makes it possible to constrain the bulk function $f$(R) by experimental/observational data.

This paper is organized as follows: in Sec. II the effective field
equations are presented in the context of $f$(\textrm{R}) models. In
Sec. III we show that brane-world $f$(\textrm{R}) effects can be 
tested by the perihelion precession of Mercury and the radar echo delay. The black hole tidal charge is then constrained by experimental/observational values. Nevertheless,   data regarding the deflection of light by the Sun are shown not be able to probe brane-world $f$(\textrm{R}) effects, being in agreement the literature for Solar system scales \cite{rindler1}. Hence, the obtainable constraint on the black hole tidal charge is led to the constraint for the DMPR black hole \cite{boemer}.
   We conclude and discuss our results in Sec. IV.

\section{Brane Field Equations for $f({\rm R})$ Gravity}

The fundamental equations for the gravitational field on the brane are quite well established. By taking the brane as the source of the gravitational field  and a 5D cosmological constant term $\Lambda _{5}$, the bulk Einstein field equations read: 
\begin{equation}
^{(5)}G_{AB}=-\Lambda _{5}\;^{(5)}g_{AB}+\kappa _{5}^{2}\;^{(5)}T_{AB },  \label{1}
\end{equation}%
where $^{(5)}G_{AB}$ denotes the 5D Einstein tensor, 
\begin{equation}
^{(5)}T_{AB}=\frac{1}{f'(R)}\left(\kappa^2_5T_{AB}^{\rm bulk}-\left(\frac
12{\rm R}f'({\rm R}) - \frac12f({\rm R}) + \Box f'({\rm R})\right)
g_{AB} + \nabla_A\nabla_B f'({\rm R})\right)\,,
\end{equation}
is the effective bulk stress tensor -- being $T_{AB}^{\rm bulk}$ the bulk stress tensor.
The brane metric $g_{\mu \nu }$ and the corresponding components of
the bulk metric $^{(5)}{g}_{\mu \nu }$ are  related by $^{(5)}{g}%
_{\mu \nu }=g_{\mu \nu }+n_{\mu }\,n_{\nu }$, where $n_{\mu }
$ is an unit vector, normal to the brane. Since $g_{55}=1$ and $g_{\mu 5}=0$ in the brane-world models here studied, namely the 5D bulk metric is given by $^{(5)}{g}%
_{AB}\,dx^{A}\,dx^{B}=g_{\mu \nu }(x^{\alpha },y)\,dx^{\mu }\,dx^{\nu
}+dy^{2}$, then  the bulk indexes effectively attain $A,B=0,1,2,3$. Moreover, $\kappa
_{5}^{2}$ stands for the 5D gravitational coupling. The brane is placed at  
$y=0$, where $y$ hereon denotes the extra dimension. 

The  matter content on the brane constitute the effective bulk stress tensor  by $
^{(5)}T_{\mu \nu }\sim S_{\mu \nu }\delta (0),$
where the delta function $\delta(0)$ is responsible for the localization on  the brane  and  $S_{\mu \nu }=-\lambda g_{\mu \nu }+\tau _{\mu \nu }$. Here $\lambda $
denotes the brane tension and $\tau _{\mu \nu }$ describes any additional
matter on the brane. The well known fine-tuning relation among the
effective 4D cosmological constant $\Lambda $ on the brane,
the bulk cosmological constant $\Lambda _{5}$, and the brane tension $%
\lambda $ is provided by  
$\Lambda =\frac{\kappa_5^2}{2}\left(\Lambda _{5}+\frac{\kappa
_{5}^{2}}{6}\lambda^2\right)$, {where the 4D coupling
constant $\kappa_4^2 = 8\pi G$ --  here $G$ denotes the 4D Newton constant -- and the  5D coupling
constant $\kappa_5^2$ are related by $\kappa_4^2 = \frac{1}{6}\lambda\kappa_5^2$}. 
The effective 4D field equations are complemented by a
set of equations obtained from the 5D  Einstein and Bianchi
equations \cite{SMS00}. On a ${\mathbb{Z}}_{2}$-symmetric brane,  induced field equations generalize the Shiromizu-Maeda-Sasaki 
procedure \cite{SMS00}, hence incorporating $f$(R) bulk effects \cite{sepangi}: 
\begin{equation}
R_{\mu \nu }-\frac{1}{2}g_{\mu \nu }R=-\Lambda g_{\mu \nu }+\kappa
_{4}^{2}\tau _{\mu \nu }+\frac{6\kappa _{4}^{2}}{\lambda }\pi _{\mu \nu
}-E_{\mu \nu }+Q_{\mu \nu }\,,  \label{555}
\end{equation}%
where $
\pi _{\mu \nu }=\frac{1}{12}\tau \tau _{\mu \nu }-\frac{1}{4}\tau _{\mu
\sigma }\tau _{\;\,\nu }^{\sigma }+\frac{1}{24}(3\tau _{\sigma \rho }\tau
^{\sigma \rho }-\tau ^{2})g_{\mu\nu}\,.$ 
Here $E_{\beta \sigma }={}^{(5)}C_{\beta \rho \sigma }^{\alpha }n_{\alpha
}n^{\rho }$, where  $^{(5)}C_{\beta \rho
\sigma }^{\alpha }$ is the bulk Weyl tensor. The term 
\begin{eqnarray}
\!\!Q_{\mu\nu}=\!\left[\frac{1}{4}f({\rm R})
\!-\!\frac{2}{5}\square f'({\rm R})\!-\!\frac{4}{15}\frac{\square f'({\rm R})}{f'({\rm R})}\!-\!\frac{{\rm R}}{10}\!\left(f'({\rm R})\!+\!\frac{3}{2}\right)\right]g_{\mu\nu}\!+\frac{2}{3}\frac{\nabla_\rho\nabla_\sigma f'({\rm R})}{f'({\rm R})}\left(\delta^\rho_{\;\mu}\delta^\sigma_{\;\nu}+n^\rho n^\sigma g_{\mu\nu}\right)\!\label{qqq}
\end{eqnarray} encompasses $f(\mathrm{R})$ bulk 
effects \cite{rengifo}. The symbol $\square$ stands for the 5D d'Alembertian, whereas for a conformally flat bulk the term $Q_{\mu\nu}$ is conserved \cite{rengifo,sepangi}.

A static spherically symmetric solution on the brane has the form 
\begin{eqnarray}
g_{\mu\nu}dx^\mu dx^\nu\!=\!-\exp({\nu(r)})dt^{2}\!+\!\exp({\lambda(r))
}dr^{2}\!+\!r^{2}d\Omega^{2},  \label{metrica}
\end{eqnarray}
where $d\Omega^{2}$ is the line element of a 2-sphere. By considering the constant Ricci
curvature scalar $\mathrm{R}$ and solving the field equations (\ref{555}) in the 
vacuum ($\tau _{\mu \nu }=0$), the topologically charged brane-world black
hole in $f(\mathrm{R})$ gravity has geometry 
 \cite{rengifo} 
\begin{eqnarray}
\exp({\nu(r)})=\exp({-\lambda(r)})=1-\frac{2GM}{c^2r}+\frac{G\beta}{4\pi\epsilon_0c^4 r^{2}}+\frac{%
\Lambda_{\mathrm{eff}}}{3}r^2,  \label{nu}
\end{eqnarray}
where $M$ is the effective mass of the black hole and%
\begin{eqnarray}  \label{lq}
\Lambda _{\mathrm{eff}}=\Lambda -\frac{Q}{4}
\end{eqnarray} (where $Q=Q_{\rho }^{\;\rho }$ as usual) 
plays the role of an effective cosmological constant on the brane, depending
upon both the brane tension and the function $f(\mathrm{R})$ as well. The parameter $%
\beta $ can be interpreted as a 5D mass parameter \cite%
{she}. It behaves as a tidal charge associated to the bulk Weyl tensor, that imparts the tidal charge stresses from the bulk to the brane \cite{Herrera-Aguilar:2015koa}. When $Q=4\Lambda$, ($\Lambda _{\mathrm{eff}}=0$), the solution
reduces to the DMPR black hole solution \cite{Da00}. For $Q=0$, the solution reduces to the topologically charged black hole
solution on the brane \cite{she}.

Regarding the particular function  $f(\mathrm{R})\sim\mathrm{R}^{n}$ in the bulk \cite{rengifo}, Eq. (\ref%
{qqq}) yields 
\begin{equation}
Q_{\mu }^{\;\nu }=\left[ \frac{\kappa _{5}^{2}\Lambda _{5}}{2}-\frac{3}{20}%
\left( \frac{10\kappa _{5}^{2}\Lambda _{5}}{5-2n}\right) ^{1/n}\right]
\delta _{\mu }^{\;\nu}. \label{q1}
\end{equation}
The brane effective cosmological constant hence  reads
\begin{eqnarray}  \label{efff}
\Lambda _{\mathrm{eff}}=\frac{\kappa _{5}^{4}\lambda ^{2}}{12}+\frac{3}{20}%
\left( \frac{10\kappa _{5}^{2}\Lambda _{5}}{5-2n}\right) ^{1/n}\,,  \label{q2}
\end{eqnarray} {leading to the fine-tuning condition when $n=1$. In fact, in this case the black hole reduces to Schwarzschild
metric with cosmological constant and tidal charge.}

On the other hand, the 4D modified models $f(R) = R +\mu^{2(n+1)}/R^n$ have been proposed in Refs. \cite{capoz,trodden}. In general, for $R^{n} \gg \mu^{2(n-1)}$, it yields  $f(R)/R\to1$. Hence there is no modification depending upon $\mu$. Notwithstanding, in the limit $R^{n} \ll \mu^{2(n-1)}$, we have $f(R)/R\to\mu^{2(n+1)}$. In this last case   scalar gravity is modified, further providing stable models \cite{od2}. In
order to agree with Solar system experiments, in Ref. \cite{vajdi} the authors obtain static spherically
symmetric solutions the case of $n = 1$, namely, for $f(R) = R+\mu/R$ theory, in both the weak and strong gravitational field regimes. From a 5D perspective, the model $%
f(\mathrm{R})=\mathrm{R}+\mu^4/\mathrm{R}$ is able to describe the positive
acceleration of the Universe \cite{rengifo,sepangi}. For a large value of $\mathrm{R}$ it gives $f(%
\mathrm{R})\sim  \mathrm{R}$, and the 5D Ricci scalar provides a
negligible modification of the usual solution. However, for small values of $%
\mathrm{R}$ gravity is modified. Possible values for $Q_{\mu \nu }$ read:
\begin{equation}
Q_{\mu }^{\;\nu }=-\frac{21\mu ^{4}}{20\left( 5\kappa _{5}^{2}\Lambda
_{5}\pm \sqrt{21\mu ^{4}+25\kappa _{5}^{4}\Lambda _{5}^{2}}\right) }\;
\delta _{\mu }^{\;\nu }\,.
\end{equation}
Hence, the effective cosmological constant on the brane takes the values%
\begin{eqnarray}
\Lambda _{\mathrm{eff}}=\Lambda +\frac{21\mu ^{4}}{20\left( 5\kappa
_{5}^{2}\Lambda _{5}\pm \sqrt{21\mu ^{4}+25\kappa _{5}^{4}\Lambda _{5}^{2}}%
\right) }\,.  \label{label}
\end{eqnarray}
In the case when $\mu \sim 0$, namely, when the
modification in $f(\mathrm{R})$ is negligible, then $\Lambda _{\mathrm{eff}}\sim\Lambda $. 

\section{Solar System Classical Tests}

The perihelion precession of Mercury, the
deflection of light by the Sun and the radar echo delay observations are
well known tests for the Schwarzschild solution of GR and
for the DMPR, the
Casadio-Fabbri-Mazzacurati, and the minimal geometric deformation in brane-world scenarios as well \cite{Casadio:2015jva}, among others. Brane-world effects in spherically symmetric
spacetimes were studied in \cite{boemer} and used in
the Solar system scrutiny. The Solar system tests can analyze properties
of topologically charged black holes in $f(\mathrm{R})$ brane-world models by constraining the parameters of $f(\mathrm{R})$ modifications and the tidal charge {proportional to} $\beta$. For topologically
charged black holes in $f(\mathrm{R})$ brane-worlds, the metric tensor
components are given by Eq. (\ref{nu}). When $\beta \rightarrow 0$ we
recover the usual general relativistic case.  In what follows we show how the Solar system tests are able to impose constraints 
on the $f$(R) bulk effects, and in particular to probe topologically
charged black holes in a $f(\mathrm{R})$ brane-world.

\subsection{The Perihelion Precession}

 The  equation of motion for a test particle under the gravitational field provided by (\ref{metrica}) reads 
\begin{eqnarray}
\dot{r}^{2}+\exp({-\lambda})\frac{L^{2}}{r^{2}}=\exp({-\lambda})\left(\frac{E^{2} 
}{c^{2}}\exp({-\nu})-1\right)\,,  \label{movi}
\end{eqnarray}
where the constants of motion $E$ and $L$, respectively, yield energy and the angular momentum conservation.
By the usual change of variables $r=1/u$ and $\dot{r}=Lu^{2}dr/d\phi $,  and 
by representing 
\begin{eqnarray}  \label{lamb}
g(u)=1-\exp({-\lambda}),
\end{eqnarray}
Eq. (\ref{movi}) reads
\begin{equation}
\left( \frac{du}{d\phi }\right) ^{2}\!\!\!+u^{2}=\!\frac{E^{2}}{%
c^{2}L^{2}}\!\exp({-\nu\! -\!\lambda})-\frac{1}{L^{2}}\exp({-\lambda})+g(u)u^{2} 
  \label{gu}
\end{equation}
and subsequently  yields 
\begin{eqnarray}
\frac{d^{2}u}{d\phi ^{2}}+u&=&\frac{1}{2}\frac{d}{du}\left(\!\frac{E^{2}}{%
c^{2}L^{2}} \exp({-\nu -\lambda})-\frac{1}{L^{2}}\exp({-\lambda})+g(u)u^{2}\right)\equiv g(u)\,.
\label{fu}
\end{eqnarray}

By denoting $\gamma(u)=\left({1-\left( {dh}/{du}\right)\vert_{u_{0}} }\right)^{1/2}$, a circular orbit $u=u_{0}$ is determined by the root of the fixed point equation $
u_{0}=h(u_{0}) $, and a deviation  is provided  by \cite{boemer} 
$
\delta =\delta _{0}\cos \left(\gamma(u)
\phi +\alpha \right)$, for $\delta _{0}$ and $\alpha$ constants. The variation of the orbital angle
 with respect to successive perihelia is
\begin{equation}
\phi =\frac{2\pi }{\gamma(u)}=\frac{ 2\pi }{%
1-\sigma },  \label{prec}
\end{equation}
\noindent where $\sigma $ is the perihelion advance, given from Eq.(\ref%
{prec}) by 
\begin{equation}  \label{masig}
\sigma \sim \frac{1}{2}\left( \frac{dh}{du}\right) _{u=u_{0}}
\end{equation}
for small values of $\left( dh/du\right) _{u=u_{0}}$. For a complete rotation the  perihelion advance  is $\delta \phi
\sim  2\pi \sigma $. 

We consider now the perihelion precession of a planet in the $f(\mathrm{R}%
)$ brane-world black hole geometry (\ref{nu}). Eq. (\ref{fu})
is thus provided by 
{\begin{eqnarray}
g(u)&=&\frac{3GMu^{2}}{c^2}-\frac{G\beta u^{3}}{2\pi\epsilon_0c^4} +\frac{GM}{c^2L^{2}}-\frac{G\beta u}{4\pi\epsilon_0c^4L^2}+\frac{%
\Lambda_{\mathrm{eff}} }{3u^{3}L^{2}}\,.
\end{eqnarray}}%
It makes $u_{0}$ to be obtained by the equation
\begin{equation}
u_0=3Mu_0^{2}-\frac{G\beta u^{3}_0}{2\pi\epsilon_0c^4}+\frac{M}{L^{2}}-\frac{G\beta u_0}{4\pi\epsilon_0c^4L^{2}}+\frac{%
\Lambda_{\mathrm{eff}} }{3u_0^{3}L^{2}}\,,
\end{equation}%
which, to first order, is approximated to $u_{0}\sim  GM/(c^{2}L^{2})$
when $\Lambda_{\mathrm{eff}}\sim 0$ and $\frac{G\beta}{4\pi\epsilon_0c^4L^{2}}\ll 1$. As $L$ is related to the orbit parameters as $L=2\pi a^{2}\sqrt{1-e^{2}} /cT$ \cite{boemer}, where $T$ denotes the period of the motion, Eq. (\ref%
{masig}) yields%
\begin{equation}
\delta \phi \!=\!\delta \phi _{GR}-\frac{\pi c^{2}}{GM}\!\!%
\left[ \frac{G\beta}{4\pi\epsilon_0c^4 a(1\!-\!e^{2})}\!+\!\Lambda_{\mathrm{eff}}
a^{3}(1-e^{2})^{3}\right],  \label{ppp}
\end{equation}%
where $\delta \phi _{GR}=6\pi GM/c^{2}a\left(1-e^{2}\right)$ is the well known Schwarzschild precession formula. Eq.  (\ref{ppp}) is consistent with the result in Ref. \cite%
{boemer}, when $\Lambda_{\mathrm{eff}}\rightarrow 0$, since  our
result incorporates $f(\mathrm{R})$ bulk effects. The above second term gives the
correction due to the nonlocal effects arising from the Weyl tensor in the
bulk \cite{boemer}.

With the observed value of the precession of Mercury perihelion given by  $%
\delta \mathring{\phi}=43.11\pm 0.21$ arcsec/century, the GR formula gives $\delta \phi _{GR}=42.98$ arcsec/century. The difference $\delta \mathring{\phi}-\delta \phi _{GR}=0.13\pm 0.21$ arcsec/century can be 
ascribed to $f(\mathrm{R})$ brane-world effects, putting stricter conditions on the results in \cite{boemer}. When this difference  results from 5D $f$(R) bulk effects on  the DMPR geometry, the bulk tidal parameter $\beta$ and the effective cosmological constant $\Lambda_{\mathrm{eff}}$ are observationally constrained by 
\begin{equation}
\!\!\bigg\vert \frac{G\beta}{4\pi\epsilon_0c^4 a(1\!-\!e^{2})}\!+\Lambda_{\mathrm{eff}}
a^{3}(1-e^{2})^{3} \bigg\vert \leq \frac{GM_\odot}{\pi c^{2}}|\delta\mathring{\phi}-\delta\phi_{GR}|.
\label{beeta}
\end{equation}
Employing the observational data  \cite{boemer}, Eq. (\ref{beeta}) provides  the parameter space
{\begin{equation}
\!\!\bigg\vert \frac{G\beta}{4\pi\epsilon_0c^4}\!+0.8\Lambda_{\mathrm{eff}}
\bigg\vert \leq (5.2 \pm 6.4)
\times 10^{4}\, \mathrm{m}^{2}.  \label{betal}
\end{equation}}

For the case $f(\mathrm{R}) = \mathrm{R}^n$, Eq. (\ref{efff})  provides the graphics for Eq. (\ref{betal}) depicted in
Fig. \ref{1} (left panel). Besides, for the case $f(\mathrm{R})=\mathrm{R}+{\mu}^{4}/\mathrm{R}$ the effective
cosmological constant is provided by Eq. (\ref{label}), and the constraint (\ref%
{betal}) is illustrated in Fig. 1 (right panel). 
\begin{figure}
  \includegraphics
  [width=0.35\textwidth]
  {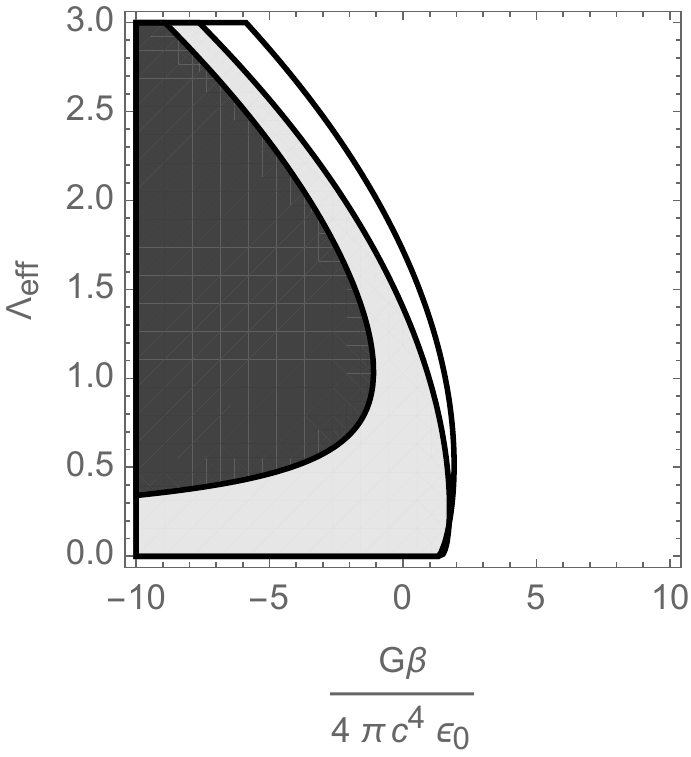} \qquad\qquad \includegraphics
  [width=0.34\textwidth]
  {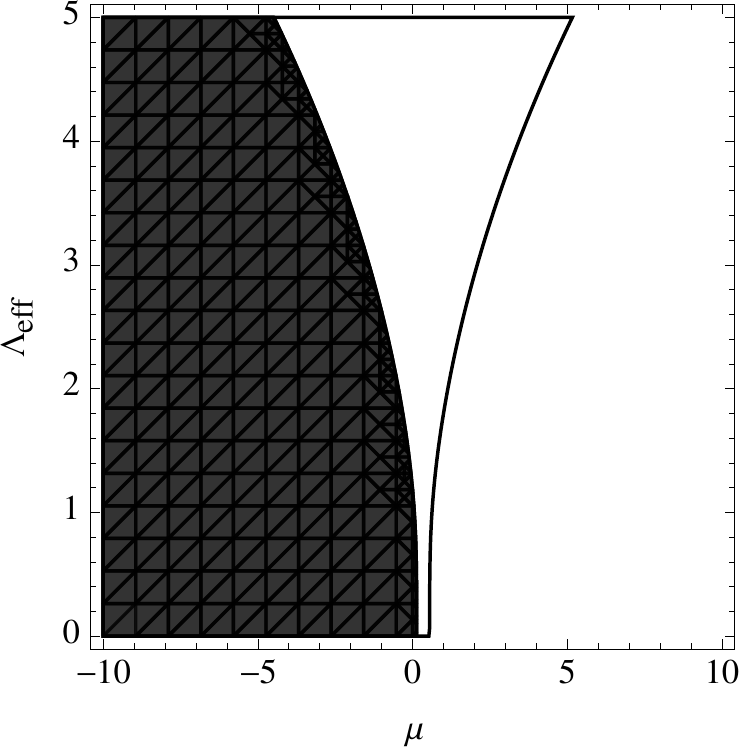}
\caption{(left panel:) graphic of the constraint (\ref{betal}) for $\lambda = 1$ in the parameter space (the tidal charge parameter $\beta$ and the effective cosmological tension $\Lambda_{\rm eff}$ are
provided in scale of $10^4$ ${\rm m}^2$) for $n=-1$ (black region), $n=3$ (the union of black and gray regions), and $n=6$ (the union of black, gray and white regions) in $f({\rm R})={\rm R}^n$ gravity; (right panel:) graphics of the constraint (\ref{betal}) in the parameter space 
in $f({\rm R})={\rm R}+(\mu^{4}/{\rm R)}$ gravity, for the positive (black region) and negative (union of black and white regions) root in (\ref{label}).}
\label{1}    
\end{figure}

\subsection{The Deflection of Light}

 A similar procedure takes into account photons on a null geodesic in the absence of external forces. The  equation of motion yields 
\begin{equation}
\left( \frac{du}{d\phi }\right) ^{2}+u^{2}=g(u)u^{2}+\frac{1}{c^{2}}\frac{
E^{2}}{L^{2}}\exp({-\nu -\lambda})\equiv p(u)\,,  \label{puu}
\end{equation} implying that 
 $\frac{d^{2}u}{d\phi ^{2}}+u=\frac{1}{2}\frac{dp(u)}{du}
$. In the lowest approximation, the solution is the line $u=%
\frac{\cos \phi }{R},$ where $R$ is the distance of the closest approach to
the mass $M$. It can be iteratively employed in the above equation, yielding
\begin{equation}
\frac{d^{2}u}{d\phi ^{2}}+u=\frac{1}{2}\frac{d}{du}\left[p\left( \frac{\cos \phi }{R}\right)\right]\ .
\label{cosine}
\end{equation}%
 The
total deflection angle of the light ray is $\delta =2\varepsilon $ \cite{boemer}.

In the case of the geometry (\ref{nu}) provided by the topologically charged
black hole in $f(\mathrm{R})$ bulk, Eq. (\ref{lamb}) leads to $g(u)=\left(
2GM/c^{2}\right) u$, resulting 
\begin{eqnarray}
p(u)&=&\frac{2GM}{c^2}u^3-\frac{G\beta u^{4}}{2\pi\epsilon_0c^4}+\frac{%
E^2}{c^2L^2}-\frac{\Lambda_{\mathrm{eff}}}{3},  \label{pudef}
\end{eqnarray}
Since the right hand side of Eq.(\ref{cosine}) 
has a derivative with respect to $u$, $f$(R) effects  encrypted in the effective cosmological constant $\Lambda_{\mathrm{eff}}$ are not perceivable. In fact,  the term $\frac{\Lambda_{\mathrm{eff}}}{3}$ in the above equation, that contains the correction induced by $f(\mathrm{R})$ effects,  does not take part on it. Hence our results are equivalent to the ones for DMPR black holes \cite{boemer,Da00}.
Clearly the total deflection of light is obtained in the same steps for the DMPR black holes \cite{boemer} 
\begin{equation}
\delta \phi=\frac{4GM}{c^2R} \left(1-\frac{3\pi \beta c^2}{16GMR%
}\right) \,,
\end{equation}
providing the constraint on the black hole charge   
{$
\left|\beta\right|\leq (7.0 \pm 27.9) \times 10^{8}\; {\rm m}^{2}\,$  \cite{boemer}.}

\subsection{Radar Echo Delay}

The radar echo delay measures the time
necessary for radar signals to travel to a planet, for instance. In fact, the time for the light to travel  between two planets that are distant from the Sun is ${\rm T}_{0}=%
\int_{-\ell_{1}}^{\ell_{2}}dx/c, $ where $\ell_{1}$ and $\ell_{2}$ are the respectively the distances from 
the planets to the Sun. On the other hand, if the light travels close to the Sun, the time
travel reads \cite{boemer} 
\begin{equation}
{\rm T}= \frac{1}{c}\int_{-\ell_{1}}^{\ell_{2}}\exp\left[\left(\lambda (r)-\nu (r)\right)/2\right]dx\,.
\end{equation}
The time difference $\delta {\rm T = T-T}_0$ is hence  given by 
\begin{eqnarray}  \label{atraso}
\delta {\rm T}&=& \!\!\frac{1}{c}%
\int_{-\ell_{1}}^{\ell_{2}}\!\!\left\{ e^{{\left[\lambda \left( \sqrt{ x^{2}+R^{2}}%
\right) -\nu \left( \sqrt{x^{2}+R^{2}}\right) \right]/2}}-1\right\} dx,\;\;\;\text{where $r=\sqrt{x^{2}+R^{2}}$.}
\end{eqnarray}

The delay can be evaluated from the integral in Eq. (\ref{atraso}). Indeed, the above  integrand is recast as:
{\begin{eqnarray}
\exp \left( \frac{\lambda }{2}-\frac{\nu }{2}\right) 
&\sim  &\left( 1-\frac{2GM}{c^2r}+\frac{G\beta}{4\pi\epsilon_0c^4 r^{2}}+\frac{%
\Lambda_{\mathrm{eff}}}{3}r^2\right) \,,
\end{eqnarray}} in a first order approximation, based upon Eq. (\ref{nu}). 
Therefore Eq.~(\ref{atraso}) reads 
\begin{eqnarray}
\delta{\rm T} &=&\frac{2GM}{c^{3}}\ln \left( \frac{\sqrt{\ell_{2}^{2}+R^{2}}+\ell_{2}}{\sqrt{\ell_{1}^{2}+
R^{2}}-\ell_{1}}\right) -\frac{G\beta }{4\pi \epsilon_0 c^5R}\left[ {\rm tan}^{-1} \left( \frac{\ell_{2}}{R}\right) +{\rm tan}^{-1}
\left( \frac{\ell_{1}}{R}\right) \right]  \nonumber \\
&&-\frac{\Lambda _{\mathrm{eff}}R^{2}}{3c}\left[ \ell_{1}\left(1+\frac{\ell_{1}^{2}%
}{3R^{2}}\right) +\ell_{2}\left(1+ \frac{\ell_{2}^{2}}{3R^{2}}\right) %
\right] \,.
\end{eqnarray}%
Using the approximations $R^{2}/\ell_{i}^{2}\ll 1$ ($i=1,2$), 
the above expression reduces to 
\begin{equation}
\delta {\rm T}\sim  \frac{2GM}{c^{3}}\ln \left( \frac{4\ell_{1}\ell_{2}}{R^{2}}\right)
-\frac{G\beta }{4c^5\epsilon_0R}-\frac{\Lambda _{\mathrm{eff}}}{9c}\left(
\ell_{1}^{3}+\ell_{2}^{3}\right) \,.  \label{bom}
\end{equation}%
This leads to the Schwarzschild radar delay $\delta {\rm T_{GR}}= \frac{2GM}{c^{3}%
}\ln \frac{4\ell_{1}\ell_{2}}{R^{2}}$ when $\beta =0$ and $\Lambda _{\mathrm{eff}%
}=0$, and to the classical test of radar echo delay for the DMPR black hole 
 when $\Lambda _{\mathrm{eff}}=0$ \cite{boemer}. The last term on the right
hand side of the above equation imposes a more strict constraint on the
class of models, in particular the ones  provided by $f(\mathrm{R})=\mathrm{R}^{n}$ and $f(\mathrm{R}%
)=\mathrm{R}+\mu ^{4}/\mathrm{R}$.

In the context of  the geometry provided by the topologically charged $f$(R) brane-world black hole metric (\ref{nu}), measurements of the frequency shift of radio photons \cite{boemer,bertoti} provide  now the following
constraint for the tidal charge parameter $\beta$ and the effective cosmological
constant: 
\begin{eqnarray}  \label{275}
\bigg\vert\frac{G\beta}{4\epsilon_0c^4}\!+\!\frac{\Lambda_{\mathrm{eff}}R_\odot}{9}%
\left(\ell_1^3 + \ell_2^3\right)\!\bigg\vert\!\lesssim (5.74\pm 6.24)\times 10^8\,{\rm m}^2\,.
\end{eqnarray}
Comparing the space of parameters 
for the DMPR black holes \cite{boemer} and for the 
topologically charged $f$(R) brane-world black hole (\ref{275}), we realize that 
$f$(R) bulk effects impose a more strict 
regime for the tidal charge $\beta$.

There is no theoretical constraint that yields the  value of $Q$ in Eq.(\ref{lq}) to be the same order of magnitude  as the 4D cosmological constant $\Lambda$ ($\sim 10^{-52}\,{\rm m}^{-2}$). Whatever the order of magnitude for the trace $Q$ of the energy-momentum tensor in Eq.(\ref{qqq}) is, it must satisfy the constraints (\ref{betal}) and (\ref{275}), accordingly. In fact,  the experimental constraint 
$
10^{32}\;{\rm m}^4\lesssim\frac{R_\odot}{9\pi}
\left(\ell_1^3 + \ell_2^3\right)\lesssim 10^{35}\;{\rm m}^4$ holds for the Solar system, and  due to the multiplication by $\frac{R_\odot}{9\pi}
\left(\ell_1^3 + \ell_2^3\right)$, the term $\Lambda_{\mathrm{eff}}$ has for the radar echo delay 
 an upper limit of $10^{-27}$ m${}^{-2}$.  It implies that this is the upper limit for the effective order of magnitude  
 of $Q$, that reflects $f$(R) bulk effects.  We shall point out our remarks in details in the next section.

\section{Concluding Remarks}

The phenomenology regarding brane-world models relies on the astronomical and
astrophysical observations at the Solar system scale. The metric for
topologically charged black holes in a $f(\mathrm{R})$ brane-world 
provides the basic theoretical tools necessary for the agreement between the theory  with the
observational/experimental results. In this context, the classical tests of
GR were considered for topologically charged black holes in
a $f(\mathrm{R})$ brane-world, and then compared to the results for the DMPR and the Schwarzschild black holes as very particular limits.

Our results encompass the DMPR black hole solution in a brane-world \cite%
{Da00}, when the parameter $\Lambda _{\mathrm{eff}}=0$. The most constrained limit
we got for the parameter $Q$ -- that encodes $f$(R) bulk effects -- came from the perihelion precession of Mercury,
and gives the constraint (\ref{betal}). These
results represent a significant restriction on tidal charge parameter \cite%
{boemer}, as the space of parameters for our model in Eq.(\ref{beeta}), illustrated in Figs.  1 for two $f$(R) models, is led to Eq.(72) of Ref.  \cite{boemer}, corresponding to the space of the parameter in the DMPR black hole. 

Although the metric (\ref{nu})  has
a Schwarzschild-AdS-like aspect when $\beta\to0$, it is completely different from the
Schwarzschild-AdS solution for such very particular case, as the effective cosmological
constant $\Lambda_{\mathrm{eff}}$ is now given by (\ref{lq}) as the sum of
the brane cosmological constant and the trace of the tensor (\ref{qqq}). For the Schwarzschild-AdS geometry, the term due to the cosmological constant does not
affect the light bending for Solar system scales \cite{rindler1}. Thus our results are in full
compliance to the literature. Indeed, for the deflection of light, Solar system observations give the same constraint  as
for the DMPR black hole \cite{boemer}.

Finally the radar echo delay, based upon 
topologically charged $f$(R) brane-world black holes,  provides a stringent constraint between the
 tidal charge parameter $\beta$ and the effective cosmological constant, provided by (\ref
{275}). 
The space of parameters (\ref{beeta}) and (\ref{275}) provides a precise range
for the trace of the tensor (\ref{qqq}) that encrypts $f$(R) effects, through the effective cosmological constant on the brane (\ref{lq}).
Moreover, since the topologically charged brane-world black hole in (\ref{nu}) presents a term containing $\Lambda_{\rm eff}$, its upper limit of $10^{-27}$ m${}^{-2}$ further provides an important constraint on the black hole geometry. It is worth to emphasize that  Eq.(\ref{lq}) further constrains  
the trace $Q=Q^\mu_{\;\,\mu}$, that arises when the  Shiromizu-Maeda-Sasaki procedure is applied to a $f$(R) bulk. Although in 4D the effect of the term $R^2$ is negligible and non observable unless the coefficient of this term is larger than $10^{61}$, the effect of higher dimensional terms is suppressed by powers of the Planck mass. The bounds on the coefficient of $R^2$ in 5D are still unknown and can be addressed, being out of the scope of our results here. We expect that it will be not so much different than that in 4D.  
Finally, nonlinear massive theories of gravity can be further analyzed in the framework here presented \cite{Cai:2013lqa}. 
\section*{Acknowledgments}

 The authors thank Prof. R.
Venegeroles and Prof. Julio M. Hoff da Silva for valuable and fruitful discussions.
A. M. K. is grateful to CAPES and ``Programa Ci\^encia sem
Fronteiras'' (CsF) for financial support. R.~d~R. thanks to FAPESP Grant No. 2015/10270-0 and CNPq Grants
 No. 473326/2013-2 and No. 303027/2012-6 for partial financial support.

\end{document}